# Method for characterizing single photon detectors in saturation regime by cw laser

Jungmi Oh,[1] Cristian Antonelli,[2] Moshe Tur,[3]
and Misha Brodsky [1,*]

[1]*AT&T Labs, 200 S. Laurel Ave., Middletown, New Jersey 07748, USA*
[2]*Department of Electrical and Information Engineering and CNISM, University of L'Aquila, L'Aquila 67040, Italy*
[3]*School of Electrical Engineering, Tel-Aviv University, Tel-Aviv 69978, Israel*
[*]*brodsky@research.att.com*

**Abstract:** We derive an analytical expression for the count probability of a single photon detector for a wide range of input optical power that includes afterpulsing effects. We confirm the validity of the expression by fitting it to the data obtained from a saturated commercial Single Photon Detector by illuminating it with a cw laser. Detector efficiency and afterpulsing probability extracted from the fits agree with the manufacture specs for low repetition frequencies.

## 1. Introduction

Photon counting by InGaAsP/InP diodes at 1550 nm wavelength was reported a quarter of century ago [1]. Since then Single Photon Detectors (SPD) have found a range of applications from quantum information experiments [2] to fiber characterization techniques [3]. To ensure proper operation of SPDs, various approaches for testing SPD have been developed. Modern advanced methods such as the interleaved bias method [4] and the time-interval analysis method [5] permit extraction of both the afterpulsing probability and the detector efficiency. However, these methods rely on complicated timing control schemes and extremely low input power levels. Thus, they require expensive equipment such as electronic time delays, pulsed lasers, high dynamic range attenuators and sensitive power meters, and might not be suitable for field applications. An end-user of recent generation portable table-top instruments could benefit from a simple method that works at relatively high power levels and requires only a conventional cw laser.

In this letter, we demonstrate a new characterization method for Single Photon Detectors that utilizes a cw laser at modestly high power ranging from –61dBm to –53dBm (for our detectors this corresponds to the average number of photon per gate, $\mu$, of 3 – 20). First, we measure probability of a count over a range of power and trigger rates for both pulsed and cw lasers. The data suggest higher afterpulsing probability at higher rates, which, unexpectedly, is accompanied by a slight decrease in detector efficiency. These tendencies are more pronounced with cw lasers than with pulsed lasers. Then, we fit the experimental results by a newly derived analytical expression using the detector efficiency and afterpulsing probability as fitting parameters. Our fits perfectly match the data within rms deviation of < 0.03%. Extracted efficiencies of 20%, and conditional afterpulsing probability in the range of 0.01 - 0.03 for 250 kHz trigger rate (device dependent) are similar to the values reported in manufacturer tests performed at the conventional low power regime of $\mu=0.1$ [4]. We found that in the cw regime the afterpulsing is somewhat stronger and decays faster with the characteristic time of 2.5-3μs, which is shorter than the 4-5μs measured for pulsed laser. This observation suggests that different processes dominate in the cw and pulsed regime. But when the trigger rates are below 200 kHz the results obtained by cw and pulsed methods are nearly identical.

## 2. Experimental setup

Figure 1 shows our experimental setup. Light from a standard telecom 1549.32nm cw laser is fed via a 3dB coupler to a variable attenuator and into a detector under test. For our experiments we used several commercial SPDs, operating in a gated mode [6-8]. To determine the effective detector gate and to compare our results with those obtained by the conventional pulsed laser method [4] we connected a 1551nm pulsed laser source (idQuantique, id300) to the other input of the 3dB coupler. The laser optical pulse width is about 30ps. We used a double stage variable attenuator with a range of 85dB. Each stage was calibrated independently at relatively high power levels of -50dBm by using a power meter. The calibration enabled us to control the power precisely at levels below the sensitivity of our power meter. Thus, the average number of photons per detector gate, $\mu$, was adjusted down to 0.02 for both cw and pulsed sources. The detector produces an electrical NIM pulse (Nuclear Instrumentation Module Standard) for each detection event. These NIM pulses were then integrated by an electronic counter.

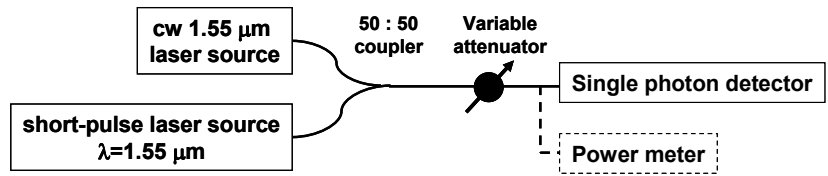

Fig. 1. The experimental setup.

In order to determine the average number of photons per detector gate in the cw regime, we measure the effective gate window of the detector, $\tau_g$. Each counting cycle, a short 1ns-wide electrical bias pulse brings the diode above the breakdown voltage. Conventionally, the maximum efficiency $\eta_0$ achieved at the peak of the bias pulse is measured with a properly aligned narrow optical pulse [4] and serves as the major spec for SPDs. On the other hand, cw light experience efficiency that changes during the bias pulse in a nonlinear fashion. To simplify calculations, we model the SPD in the cw regime as operating at its maximal efficiency value $\eta_0$ for an effective gate window $\tau_g$. To assess $\tau_g$ experimentally, we first measure the probability of a count $P_c$ over a wide power range of each cw and pulsed laser. The repetition rate of the detector is $R=100kHz$ in both cases, and the pulsed laser is synchronized with the detector. Then we pick the values of the cw laser power $P_{cw}$ and the average pulsed laser power $P_p$ that equates the corresponding count probabilities $P_c(P_{cw}) = P_c(P_p)$. Finally the gate window can be calculated as: $\tau_g = 10^{(P_p - P_{cw} - 10\log_{10}(R))/10}$, where the power values $P_p$ and $P_{cw}$ are in dBm. When averaged over the entire power range this procedure results in $\tau_g$ in a range of 0.49ns – 0.59ns for different detectors.

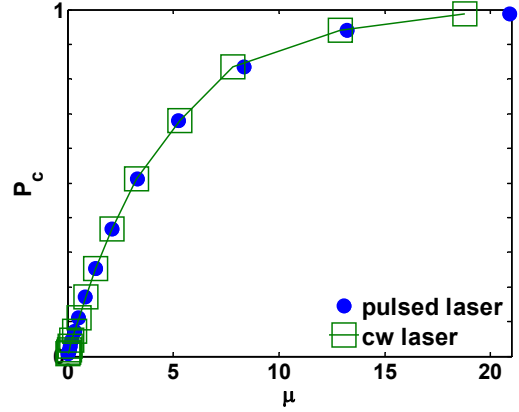

Fig. 2. Measured count probability $P_c$ as a function of the average number of photons per gate $\mu$ for pulsed (●) and cw (□) laser. Here $\tau_g = 0.49ns$.

Figure 2 plots measured count probabilities $P_c(\mu)$ as a function of the average number of photons per gate $\mu_{cw} = \frac{10^{(P_{cw}/10)}}{1000 \times h \times \nu} \times \tau_g$ and $\mu_p = \frac{10^{(P_p/10)}}{1000 \times h \times \nu \times R}$, where $h$ is the Planck constant and $\nu$ is optical frequency. Here the value $\tau_g = 0.49ns$ equates the measured count probabilities in the cw and the pulsed. Although here we determined the effective gate window $\tau_g$ by ourselves, it could preferably be provided by the manufacturer as one of the instrument's specs.

### 3. Derivation of the detector count probability

The dependence shown in Fig. 2 saturates because avalanche SPDs are not sensitive to the number of registered photons. For the Poissonian photon statistics, this saturation was first described by a simple expression in [1],

$$P_c = 1 - (1 - P_{dc}) \times \exp(-\mu\eta) \qquad (1)$$

Here $P_{dc}$ is the probability of a dark count, $\mu$ is the average number of photons per SPD gate and $\eta$ is the detector efficiency. However, we find that our data deviates from such dependence in two ways. First, we find that the measured probability is somewhat higher at lower $\mu$ values, which suggests the presence of afterpulses [7-8]. Second, the entire curve seems to be tilted toward higher $\mu$ indicating a slight decrease in efficiency. These two effects become more pronounced with higher repetition rates. To account for these deviations we modify the expression for the count probability to inlcude the afterpulsing effect.

We characterize the afterpulsing effect by the conditional probability $Q(t)$ of having an avalanche for the detector gate being open after a time interval $t$ since the successful detection event. We further assume that this conditional probability decays exponentially with a time constant $\tau$: $Q(t) = Q_0 \exp(-t/\tau)$ [9]. Then $Q_n$ is the conditional probability of the avalanche occurrence after exactly $n$ clock cycles in the detector that is gated with the trigger rate $R$:

$$Q_n = Q_0 \exp(-n/R\tau) \quad (2)$$

Within this model the overall probability of a registered event $P_c$ can be found to be

$$P_c = \frac{2 - e^{1/R\tau} - (1 - P_{dc})(1 - P_{ph})(1 - Q_0)}{2} + \frac{1}{2}\sqrt{\left[2 - e^{1/R\tau} - (1 - P_{dc})(1 - P_{ph})(1 - Q_0)\right]^2 + 4\left(e^{1/R\tau} - 1\right)\left[1 - (1 - P_{dc})(1 - P_{ph})\right]} \quad (3)$$

Here $P_{ph}$ would be probability of detecting photons generated by the source under scrutiny in the absence of both dark counts and afterpulses. For a Poissonian source of single photons, such as attenuated coherent laser light, $P_{ph} = 1 - \exp(-\mu\eta)$. As an aside, we would like to mention that if a low-power Poissonian source of photon pairs is available instead of attenuated laser light, then Eq. (3) still can be used, but with $P_{ph} = 1 - \exp(-\mu\eta(2-\eta))$.

The derivation of Eq. (3) is accomplished in two steps. The first step consists of evaluating the marginal probability of an afterpulse $P_{ap}$:

$$P_{ap}^{(n)} = P_c^{(n-1)}Q_1 + \left(1 - P_c^{(n-1)}\right)P_c^{(n-2)}Q_2 + \left(1 - P_c^{(n-1)}\right)\left(1 - P_c^{(n-2)}\right)P_c^{(n-3)}Q_3 + \ldots \quad (4)$$

where the superscript $n$ denotes that the corresponding probability is evaluated at the $n$-th time interval. Note that $Q_k$ is a conditional probability of an afterpulse given that no counts occur during the time $t = k/R$. Substitution of Eq. (2) into Eq. (4) yields:

$$P_{ap}^{(n)} = P_c^{(n-1)}Q_1 + P_{ap}^{(n-1)}\left(1 - P_c^{(n-1)}\right)Q_1/Q_0 \quad (5)$$

After a some large number of trigger cycles the probabilities $P_{ap}^{(n)}$ and $P_c^{(n)}$ reach their stationary values and can be replaced by $P_{ap}$ and $P_c$, respectively, in Eq. (5), which hence can be solved for $P_{ap}$, yielding

$$P_{ap} = \frac{P_c Q_1}{1 - (1 - P_c)Q_1/Q_0} \quad (6)$$

Finally, the second step of our derivation consists of extracting $P_c$ from $P_c = 1 - (1 - P_{dc})(1 - P_{ph})(1 - P_{ap})$, where $P_{ap}$ is given by Eq. (6). Note that for $Q_0 \to 0$ or $\tau \to 0$ (that is $Q_1 \to 0$) Eq. (6) yields $P_{ap} = 0$, hence $P_c$, as evaluated from $P_c = 1 - (1 - P_{dc})(1 - P_{ph})$, reduces to the simple form given by Eq. (1).

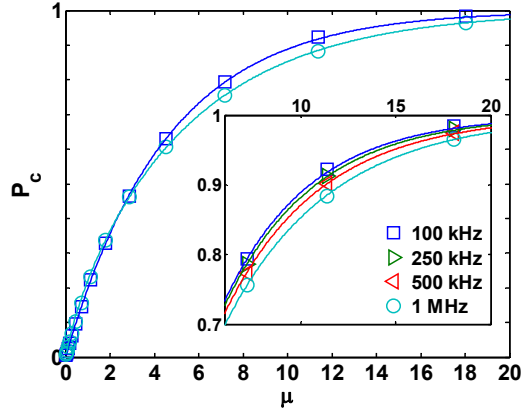

Fig. 3. Count probability $P_c$ versus average number of photons per gate $\mu_{cw} = 0.1 - 20$ for cw laser at different trigger rates of 100 kHz (□) and 1 MHz(○). Inset: a zoom in with more trigger rates. Data is shown by symbols, lines are fits.

## 4. Experimental results

Figure 3 shows the measured count probability for cw laser for the average incoming photon number $\mu_{cw} = \dfrac{10^{(P_{cw}/10)}}{1000 \times h \times f} \times \tau_g$ ranging from 0.1 to 20. The data taken at each of the four different trigger rates is plotted in different symbols. Our calculation presented above shows that for each trigger rate $R$ the experimental dependence of $P_c$ on $\mu_{cw}$ can be described by the expression in Eq. (3), computed with the corresponding values of $R$. This allows us to extract the efficiency for each trigger rate $\eta(R)$, the afterpulsing constant $Q_o$, and the detrap time $\tau$ by fitting Eq. (3) to the data.

We first measure the rate dependent dark count probability $P_{dc}$. Then we perform a simultaneous fit to all four experimentally obtained dependencies of the count probability on the average photon number. By inserting the resulting parameters back into Eq. (3) we plot lines in Fig. 3. The high quality of the fits can be clearly seen. In fact, by using only a limited range of $\mu_{cw} = 3 - 20$ (corresponding to easily measurable interval of $-61 dBm \leq P_{cw} \leq -53\ dBm$ ) we obtained a nearly perfect match to our data within rms deviation of $3 \times 10^{-4}$. Thus we conclude that the basic detector parameters can be extracted from measured photon counts at a relatively high number of incoming photons.

Additionally, we apply our high power characterization method for both pulsed and cw lasers. The extracted efficiency is shown on the top panel of Fig. 4 as squares (cw source) and triangles (pulsed source). For low repetition rates $R = 100kHz$ and $R = 250kHz$ the two methods produce nearly identical results. At $R = 250kHz$ the measured efficiency is 21%, which is sufficiently close to the spec value of 20% at that frequency. Our data indicate that the efficiency drops somewhat with repetition rate. However, this drop is rather dramatic for the cw regime and results in 2% deviation from the value of 19.7 %, measured in the pulsed regime at $R = 1MHz$. The observed trends in the efficiency are also reflected in the afterpulsing probabilities. The latter is illustrated in the bottom panel as $Q(t) = Q_0\ exp\left(-t/\tau\right)$, with the constant $Q_0$ and detrap time $\tau$ extracted by the fits. Again we obtain for the two regimes very similar results for times longer than $4 - 5\mu s$, corresponding to rates less than $200 - 250kHz$. For small time values afterpulsing for the cw regime grows significantly. In fact, the cw detrap time $\tau_{cw} \approx 2.5 - 3\mu s$ appears to be somewhat smaller than the values obtained for the pulsed regime $\tau_{pulsed} \approx 4-5\mu s$, which are more consistent with those reported by the manufacturer [7-8]. We speculate that the relaxation in the cw regime might be a combination of several processes beyond afterpulsing, with the other processes such as the charge persistence or twilight effect being dominant for small time values in the cw regime [10-11].

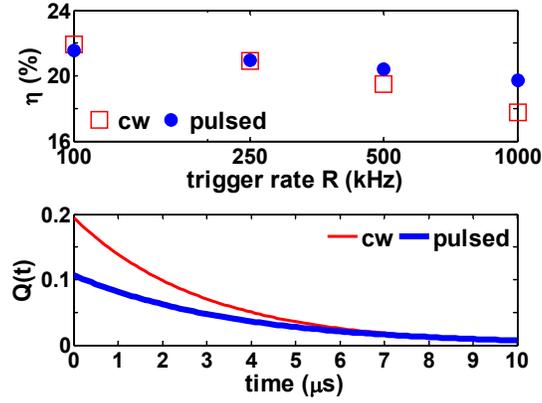

Fig. 4. Top: extracted detector efficiency $\eta$ at different trigger rates $R$ for cw (□) and pulsed (●) lasers. Bottom: conditional afterpulsing probability for cw (red thin line) and pulsed (blue thick line) measurements.

## 5. Conclusion

We presented a new method of extracting the SPD efficiency and the afterpulsing probability. The method consists of measuring photon counts at relatively high cw power at several repetition rates and fitting the data with a newly derived analytical expression. We show that the method produces good results for repetition rates below $250 kHz$. Such a simple and reliable method could potentially help a service provider with maintenance and configuration of QKD systems.